\documentstyle[11pt,mrs2003,epsfig]{article}
\begin{document} 

\def\Ok{$\Omega_{\rm k}\ $}
\def\Om{$\Omega_{\rm M}\ $}
\def\Ol{$\Omega_{\rm \Lambda}\ $}
\def\Or{$\Omega_{\rm R}\ $}
\def\Ogn{$\Omega_{\rm {\gamma}{\nu}}\ $}
\def\Neff{$N_{\nu ,eff}\ $}
\def\lcdm{$\Lambda$CDM }
\def\rhocrit{\rho_{\rm crit}}

\def\himpc{$h^{-1}Mpc\ $}

\def\etal{{\it et al.}~}
\def\eg{{\it e.g.},~}
\def\ie{{\it i.e.},~}

\def\4he{$^4$He}
\def\d{D$/$H}
\def\7li{$^7$Li}
\def\8Be{$^8$Be}
\def\wq{w_{\rm Q}}
\def\Mkev{M_{\rm keV}}
\def\dd{\textrm{d}}

\def\Msun{$M_{\odot}$,~}
\def\Msol{M_{\odot}}
\def\V{V(\phi)}
\def\Mvir{M_{\rm vir}}
\def\Rvir{R_{\rm vir}}
\def\cvir{c_{\rm v}}
\def\Vvir{V_{\rm vir}}
\def\Vmax{V_{\rm max}}
\def\dvir{\Delta_{\rm vir}}
\def\rs{r_{\rm s}}
\def\dv2{\Delta_{\rm V/2}}

\def\lsim{\lower0.6ex\vbox{\hbox{$ \buildrel{\textstyle <}\over{\sim}\ $}}}
\def\gsim{\lower0.6ex\vbox{\hbox{$ \buildrel{\textstyle >}\over{\sim}\ $}}}

\def\Rv2{r_{\rm V/2}}
\def\Mcoll{M_*(z_c)}
\def\rhoc{\tilde\rho_{\rm s}}
\def\Mchi{{\rm{M_\chi}}}
\def\rmin{r_{\rm min}}

\def\hMpc{h^{-1}{\rm Mpc}}
\def\Msun{\rm M_{\odot}}
\def\hMsun{h^{-1}M_{\odot}{\ }}
\def\M9{M_9}

\def\Mvir{M_{\rm v}{\ }}
\def\Vvir{V_{\rm v}{\ }}
\def\Vmax{V_{\rm m}}
\def\Rvir{R_{\rm v}}
\def\Cvir{c_{\rm v}}
\def\Mv{M_{\rm v}}
\def\Vv{V_{\rm v}}
\def\Vmax{V_{\rm max}}
\def\Rv{R_{\rm v}}
\def\Cv{c_{\rm v}}
\def\Nv{N_{\rm v}}

\def\rs{r_{\rm s}}

\def\photon2{\gamma \gamma}
\def\Zgamma{\gamma Z^0}

\def\LCDM{$\Lambda$CDM}
\def\Omegan{$\Omega_0$}
\def\kms{\ {\rm km\,s^{-1}}}
\def\hmpc{\ \textrm{h}^{-1}{\rm Mpc}}
\def\hkpc{\ \textrm{h}^{-1}{\rm kpc}}
\def\ihmpc{\ \textrm{h}~{\rm Mpc^{-1}}}
\def\himsun{\textrm{h}^{-1}\textrm{M}_{\odot}}
\def\hikpc{\textrm{h}^{-1}\textrm{M}_{\odot}}
\def\kmps{\textrm{km}\ \textrm{s}^{-1}}

\def\equ#1{equation~\ref{eq:#1}}
\def\Fig#1{Figure~\ref{fig:#1}}
\def\Deltavir{\Delta_{\rm vir}}
\def\LCDM{\Lambda {\rm CDM}}
\def\cv{c_{\rm v}}
\def\GeV{{\rm GeV}}
\def\cm{{\rm cm}}
\def\s{{\rm s}}
\def\sr{{\rm sr}}
\def\Aeff{{\cal{A}}_{\rm eff}}
\def\LOS{{\rm LOS}}
\def\TeV{{\rm TeV}}
\def\texp{t_{\rm exp}}
\def\ymax{y_{\rm max}}
\def\Eth{E_{\rm th}}
\def\Ns{N_{\rm s}}
\def\NB{N_{\rm B}}
\def\OmegaMatter{\Omega_{\rm M}}
\def\OmegaLambda{\Omega_{\rm \Lambda}}
\def\rc{r_{\rm c}}
\def\rs{r_{\rm s}}
\def\rtilde{\tilde{r}}
\def\RSUSY{\Re_{\rm Susy}}
\def\LHALO{{\cal{L}_{\rm halo}}}
\def\rtildec{ {\tilde{r}}_{\rm c}}
\def\rtildeco{ {\tilde{r}}_{\rm c,0}}
\def\rhoc{\rho_{\rm c}}
\def\rhos{\rho_{\rm s}}
\def\Mmin{M_{\rm min}}
\def\L{{\cal{L}}}

\title{Gamma-rays From Neutralino Annihilation in Milky Way Substructure:  What Can We Learn?} 

\author{Savvas M. Koushiappas}
\affil{Department of Physics, The Ohio State University, Columbus, OH
43210, USA\\smkoush@mps.ohio-state.edu}
\author{Andrew R. Zentner}
\affil{Department of Physics, The Ohio State University, Columbus, OH 43210, USA}
\affil{Present Address: Center for Cosmological Physics, The University of Chicago,\\ 
Chicago, IL 60637, USA}
\author{Terrence P.  Walker} 
\affil{{Department  of Physics, The Ohio State  University, 
Columbus, OH 43210, USA}}
\affil{Department of Astronomy, The Ohio State University, 
Columbus, OH 43210, USA}

\begin{abstract} 

We  estimate   the  probability  of  detecting   gamma-rays  from  the
annihilation  of neutralino  dark matter  in the  substructure  of the
Milky Way.  We characterize  substructure statistically based on Monte
Carlo realizations  of the  formation of a  Milky Way-like  halo using
semi-analytic  methods  that   have  been  calibrated  against  N-body
simulations.  We find  that it may be possible  for the upcoming GLAST
and VERITAS experiments, working in concert, to detect gamma-rays from
dark  matter substructure  if $\Mchi  \lsim  100 \,  \GeV$, while  for
$\Mchi  \gsim  500  \,  \GeV$  such a  detection  seems  unlikely.  We
investigate the effects of  the underlying cosmological model and find
that the probability of detection is sensitive to the primordial power
spectrum of density fluctuations  on small (galactic and sub-galactic)
scales.  We conclude that the lack of such a detection reveals little
about the supersymmetric parameter space due to the uncertainties 
associated with the properties of substructure and cosmological 
parameters.

\end{abstract} 
 
\section{Introduction} 

In  the  currently popular  $\Lambda$CDM  cosmological model,  the  Universe  is
composed   of  $\sim   4$\%   baryonic  matter and $\sim  26$\%   cold,
collisionless  dark matter  (CDM), and is made flat  by  a cosmological
constant ($\Lambda$)  \cite{wmap}.  The growth of  structure is seeded
by  a   nearly  scale-invariant  spectrum   of  density  fluctuations,
supposedly generated during an  early epoch of inflation.  Within this
framework,   structure  forms   hierarchically,  with   small  objects
collapsing first and subsequently  merging into larger structures over
time.  This paradigm for  structure formation predicts the presence of
a  large number of  self-bound subhalos  within Milky  Way-sized halos
(e.g., \cite{dsp}) and it is possible that the these substructures may
give rise  to a gamma-ray signal  due to annihilations  of dark matter
particles in  their dense inner regions  \cite{past_studies}.  This is
based on the  assumption that the dark matter  is a weakly-interacting
massive particle  (WIMP) that annihilates  into photons.  Such  a WIMP
candidate  is provided by  supersymmetry (SUSY).  In the  most popular
SUSY models,  R-parity conservation guarantees that  the lightest SUSY
particle  (LSP)  is stable.   Additionally,  a  large  region of  SUSY
parameter space provides an LSP  with the requisite relic abundance to
serve as the CDM.  In the constrained minimal supersymmetric extension
to the standard model (MSSM),  this particle is typically the lightest
neutralino, or  lightest mass eigenstate  formed from the  two CP-even
Higgsinos, the $W^3$ino and the Bino.

In this Proceeding, we  explore the idea that neutralino annihilations
in  Milky  Way  (MW)  substructure  may teach  us  about  SUSY  and/or
structure formation.  In particular,  we estimate the probability that
the  gamma-ray signal  from neutralino  annihilations  in substructure
will  be  detected by  the  upcoming  GLAST  and VERITAS  experiments,
assuming  the  the majority of the CDM  is  in  the  form  of  neutralinos.   
Further,  we investigate the type of information that may be gleaned from such 
a gamma-ray detection, or lack thereof.   The results  that we summarize  
here are based  on the work of Ref. \cite{KZW03}, to which we refer the 
reader for details.

\section{\label{sec:substructure}Milky Way Substructure}

To begin with, we describe  the matter density profiles of halos using
the result of  Navarro, Frenk, and White (NFW)  \cite{NFW}: $\rho(r) =
\rhos (r/r_s)^{-1} (1+r/r_s)^{-2}$.  This  description of CDM halos is
supported  by  the most  recent  numerical  studies \cite{power}.   To
estimate  the properties  of  substructure  in the  MW,  we adopt  the
simple,  semi-analytic  model  described  in  Ref. \cite{ZB03}.   We  first
generate Monte  Carlo realizations  of the merger  history of  a Milky
Way-sized  host  halo  using  the extended  Press-Schechter  formalism
\cite{EPS}.   We then  track  the orbit  of  the subhalo  in the  host
potential  in  order  to  determine  whether or  not  the  subhalo  is
destroyed by  tidal forces  and to estimate  its final  position. This
model produces substructure  radial distributions, mass functions, and
velocity functions that are  in approximate agreement with the results
of  high-resolution  N-body simulations.   This  method  allows us  to
account  approximately  for  known  correlations between  the  density
structure and collapse histories of subhalos, to model substructure in
a simple  way that  is inherently free  of resolution effects,  and to
generate  statistically significant  results  for a  variety of  input
parameters by examining a large number of realizations of MW-like host
halos.

\section{The Gamma-ray Signal From Substructure}

We  calculate  the  number   of  gamma-ray  photons  originating  from
neutralino  annihilations  in  the  central  regions  of  subhalos  by
assuming the  best-case-scenario for detection.  We choose  to fix the
annihilation   cross  section  into   all  intermediate   states  that
subsequently  decay  and/or hadronize  to  yield  photons to  $\langle
\sigma |v| \rangle_{\rm h} = 5 \times 10^{-26} \cm^3 {\rm s}^{-1}$ and
the  annihilation cross  section  into the  2-photon and  $Z^0$-photon
final states to be $\langle \sigma |v| \rangle_{\gamma \gamma , \gamma
Z^0} = 10^{-28} \cm^3  {\rm s}^{-1}$.  These values are representative
of the  maximum achievable  cross sections within  the context  of the
constrained MSSM.   We include the  contributions from the  cosmic ray
electron  \cite{nishimura},   hadron  \cite{ryan},  and   the  diffuse
gamma-ray backgrounds \cite{sreekumar} in our calculation of competing
backgrounds.   We adopt  a  liberal  definition of  a  detection at  a
significance of $S > 3$  and note that due to detector specifications,
the  significance is  a function  of threshold  energy  and neutralino
mass.  In  accord with our  strategy of optimizing the  likelihood for
detecting the  gamma-ray signal  from substructure, we  concentrate on
observations  at a  threshold  energy and  neutralino  mass where  the
significance   is  maximized.    Using  the   specifications   of  the
atmospheric           {\v{C}}erenkov          telescope          (ACT)
VERITAS\footnote{http://veritas.sao.arizona.edu},  we  find  that  the
significance is maximized at a neutralino mass of $\Mchi \simeq 500 \,
\GeV$, with a threshold energy of  $E_{\rm th} \simeq 50 \, \GeV$.  We
take  GLAST\footnote{http://glast.gsfc.nasa.gov} as  an  example of  a
space-based detector. In this case, the significance is maximized at a
neutralino mass  just above experimental  bounds, $\Mchi \simeq  40 \,
\GeV$, and at a threshold $E_{\rm th} \simeq 4 \, \GeV$.

\section{\label{sec:results}Results}

\begin{figure}  
\label{fig:fig1}
\vspace*{1.25cm}
\begin{center}
\begin{tabular}{cc}
\epsfig{figure=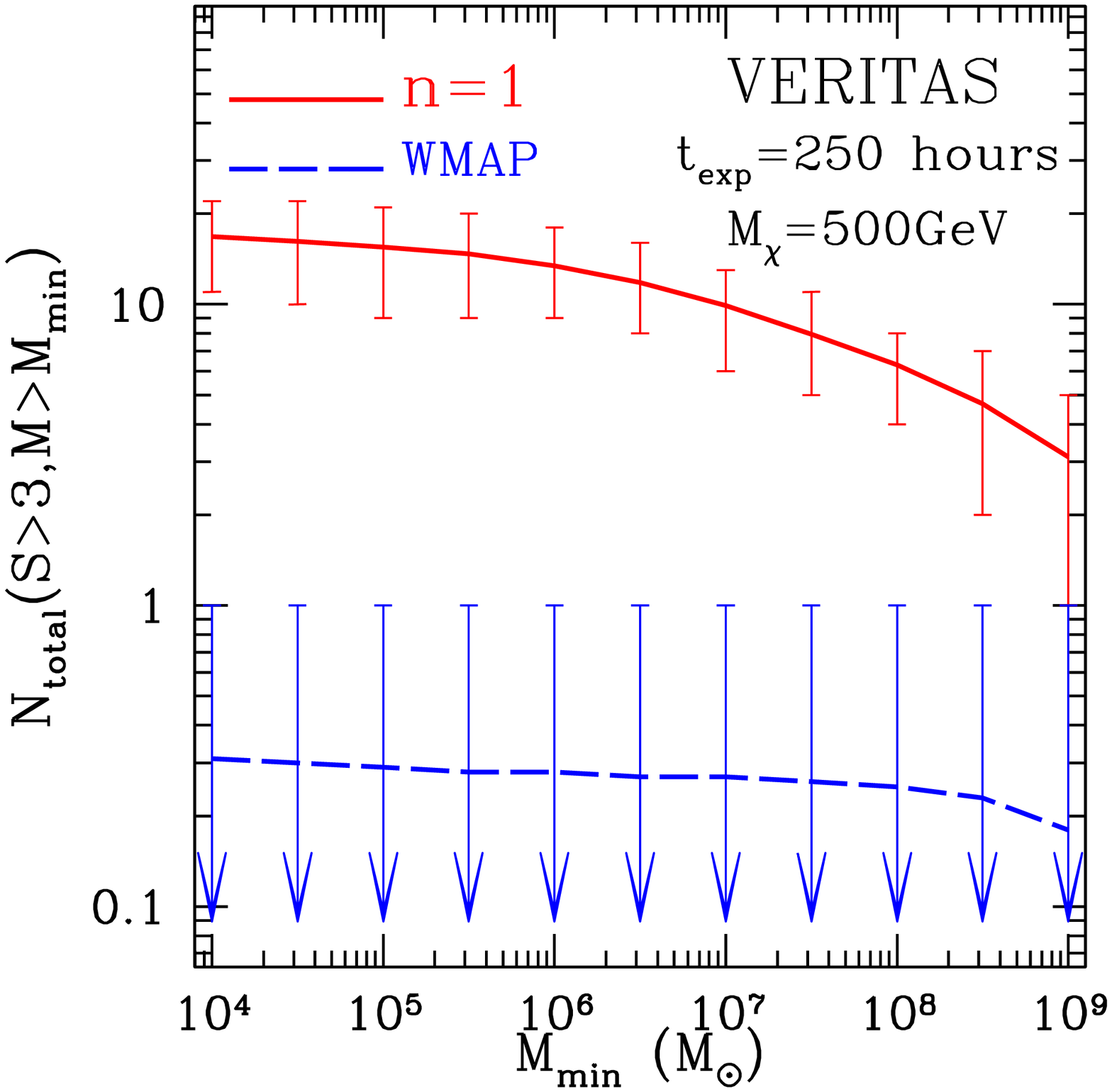, width=6.5cm} \epsfig{figure=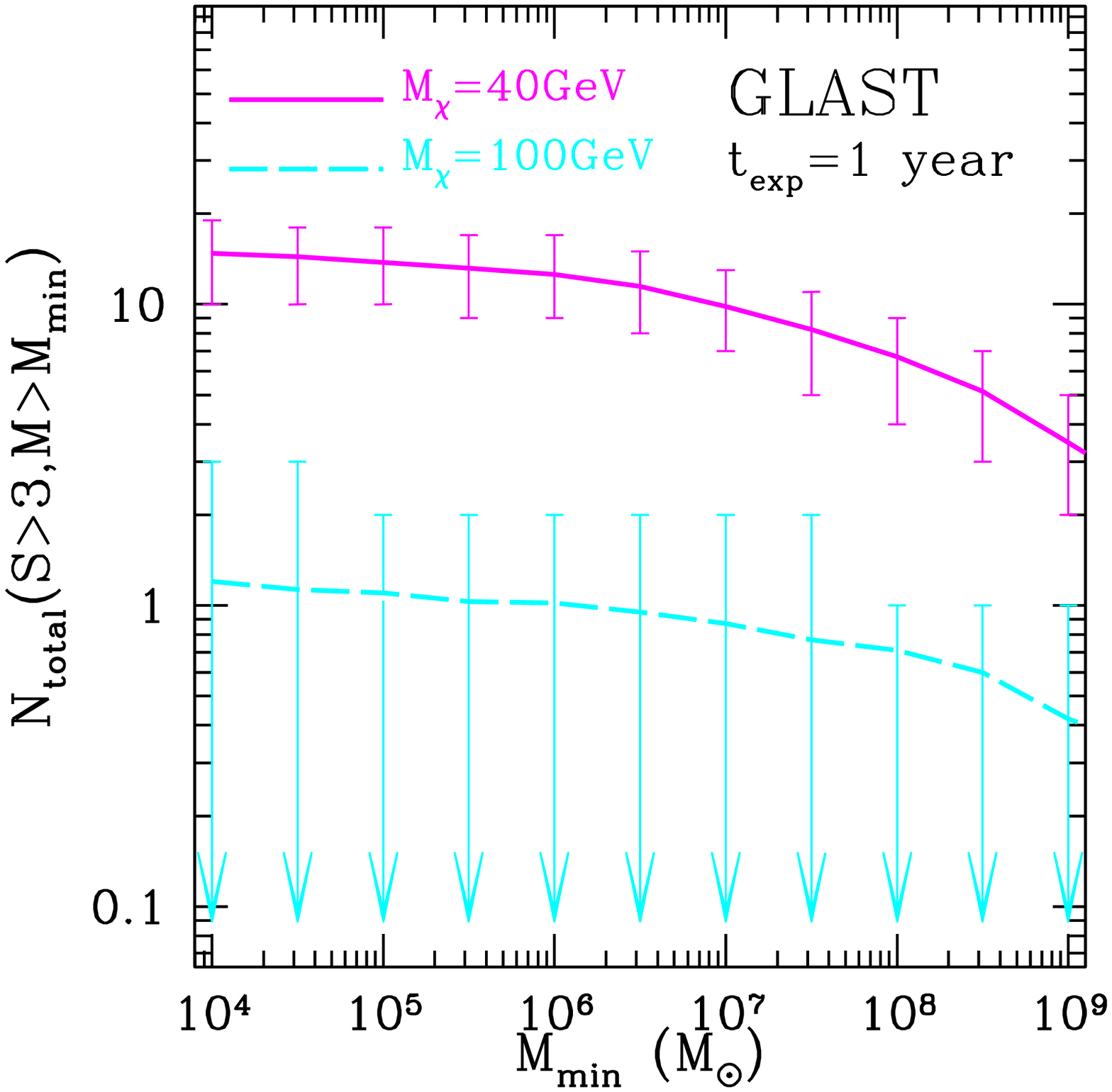,width=6.5cm}
\end{tabular}
\end{center}
\vspace*{0.25cm}
\caption{Left:  The cumulative number  of subhalos  on the  entire sky
with mass  $M \ge \Mmin$ that are  detectable at $S \ge  3$ by VERITAS
after an exposure time of 250  hours.  The neutralino mass is $\Mchi =
500 \, \GeV$ and $\Eth = 50 \, \GeV$, yielding the highest probability
for  detection. The  solid line  represents  the mean  over 100  model
realizations  in the $\LCDM$  cosmological model  with scale-invariant
primordial power  spectrum. The dashed  line represents the  mean over
100 realizations  in a  $\LCDM$ model with  a running power  law power
spectrum, $\textrm{d} n /\textrm{d} \ln k = -0.03$ 
with $n(k=0.05 \textrm{ Mpc}^{-1}) \simeq 0.93$ 
and $\sigma_8 \simeq 0.84$.  In both cases, the  error bars correspond to the 64\% range
of the  predictions (symmetric about the median). The down arrows  indicate that more 
than 18\% of the realizations had zero visible halos in the corresponding mass bin.
Right: The cumulative number of  visible subhalos detectable at $S \ge
3$, with mass  $M \ge \Mmin$, after a one year  exposure with GLAST in
standard $\LCDM$. The  threshold energy is $\Eth =  3 \, \GeV$.  Solid
lines represent the mean over  100 model realizations for a neutralino
with  mass  $\Mchi =  40  \, \GeV$,  while  dashed  lines represent  a
neutralino with $\Mchi=100\, \GeV$.  The error bars are as in the left
panel.}
\end{figure}

Our first results deal with  the likelihood of a detection by the VERITAS 
instrument.  Figure \ref{fig:fig1} shows results for the most optimistic case for
VERITAS, namely, $M_{\chi}  = 500$ GeV observed  above a threshold  of $\Eth =
50$ GeV.  We  have assumed a generous exposure time  of $t_{\rm exp} =
250$ hours.  In Fig. \ref{fig:fig1},  we show the cumulative number of
visible subhalos in  the entire sky as a function  of an adopted lower
mass cut-off, $\Mmin$.  In practice,  there is likely a cut-off in CDM
substructure at  some low  mass, well beyond  the regime  where N-body
simulations  can probe.   Our results  indicate that  reducing $M_{\rm
min}$, adding more  low-mass subhalos, does not necessarily  lead to a
dramatic  increase  in the  number  of  visible  subhalos.  Our  model
suggests  that the  number of  detectable subhalos  per  mass interval
scales as  $\textrm{d} N_{\rm total}/\textrm{d} \ln  M \sim M^{-0.02}$
at low mass.   We find that with $M_{\rm min}  = 10^4$ M$_{\odot}$, we
expect $N_{\rm  total} \sim 17$  detectable subhalos, with  fewer than
$N_{\rm total}  \sim 25$  detectable at $95\%$.   This means  that, on
average, an  ACT like VERITAS will  have to survey $\sim  1/20$ of the
sky to find {\em one} subhalo.  Considering the small field of view of
VERITAS and the fact that ACTs can, on average, observe $\sim 6$ hours
per night,  the prospects for detection must rely heavily  on serendipity,
even in  the best of  circumstances.  Note that for  neutralino masses
higher or lower than $M_\chi \sim 500$ GeV, there are fewer detectable
subhalos.

If we  are to  learn about SUSY  and/or structure formation  from such
experiments, we  must investigate the sensitivity of  these results to
cosmological  parameters.   In  Figure  \ref{fig:fig1},  we  show  the
results of a  calculation based on a $\LCDM$  model with a nonstandard
power  spectrum.  We  show a  model with  a running  power  law index,
$\textrm{d}n/\textrm{d} \ln  k = -  0.03$, as suggested by  the recent
analysis of the WMAP team\cite{wmap}. In this case, the mean number of
visible subhalos is reduced by  more than an order of magnitude.  This
can be explained by examining the effects of reduced small-scale power
on the  properties of  substructure populations \cite{ZB03}  and shows
that predictions  for the gamma-ray signal from  WIMP annihilations in
substructure are sensitive the power spectrum on sub-galactic scales.

We  now turn  our attention  to  the space-based  GLAST detector.   In
Figure \ref{fig:fig1},  we show the  number of subhalos that  would be
detectable after a year long exposure with GLAST.  The most optimistic
number   of  detectable   subhalos   is  $N_{\rm   total}  \sim   14$,
corresponding to roughly  {\em two} subhalos per GLAST  field of view;
however, one must be cautious.  Consider the energy scales involved in
this calculation.   The optimum neutralino mass for  a GLAST detection
is at the lower limit of current experimental searches, $M_{\chi} \sim
40$  GeV.  Increasing  the neutralino  mass results  in  fewer visible
halos  due  to the  limited  effective area  of  GLAST  and a  rapidly
decreasing subhalo luminosity.

\begin{figure} 
\label{fig:fig2}
\vspace*{1.25cm}
\begin{center}
\begin{tabular}{cc}
\epsfig{figure=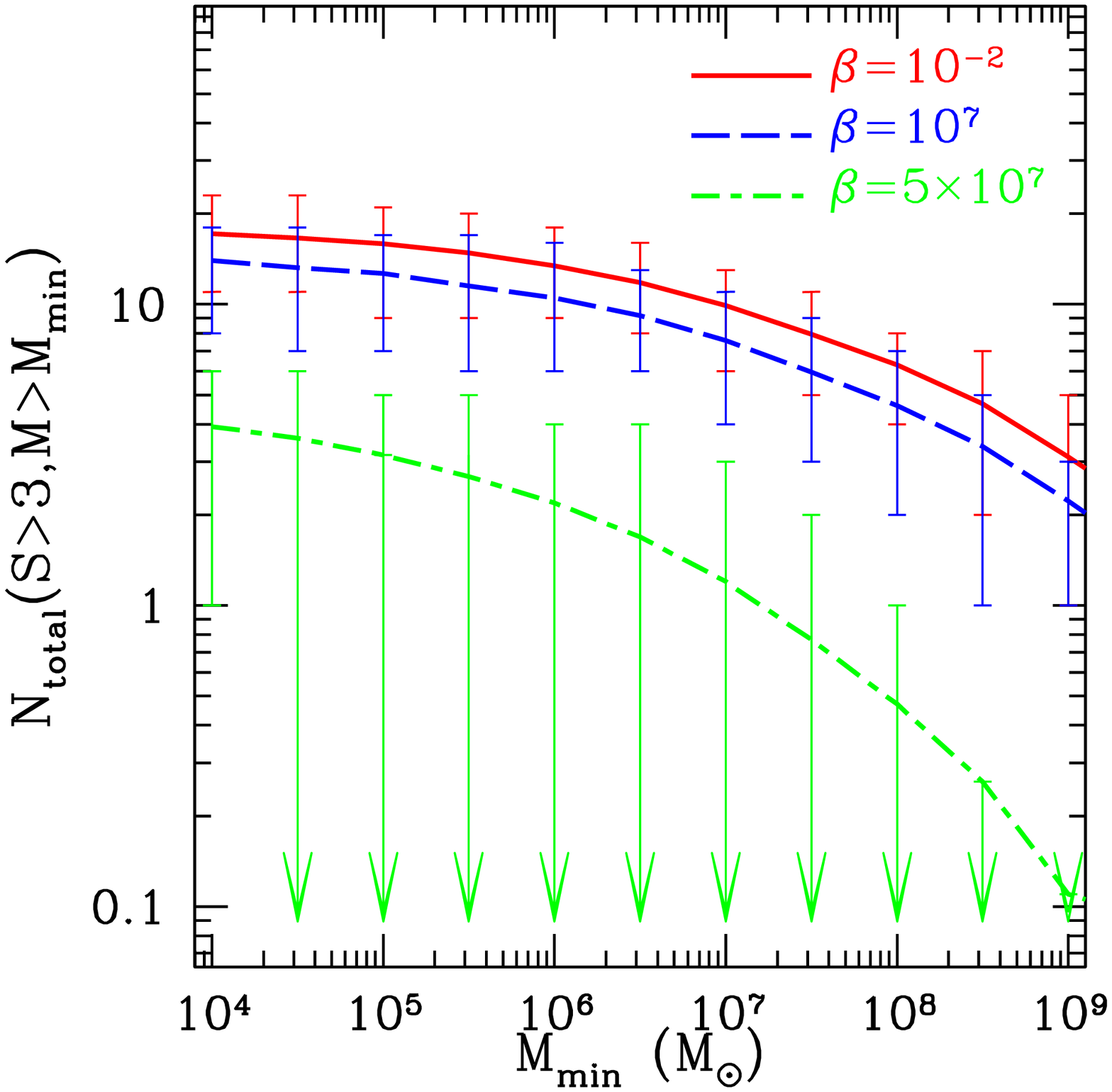,width=6.5cm}
\epsfig{figure=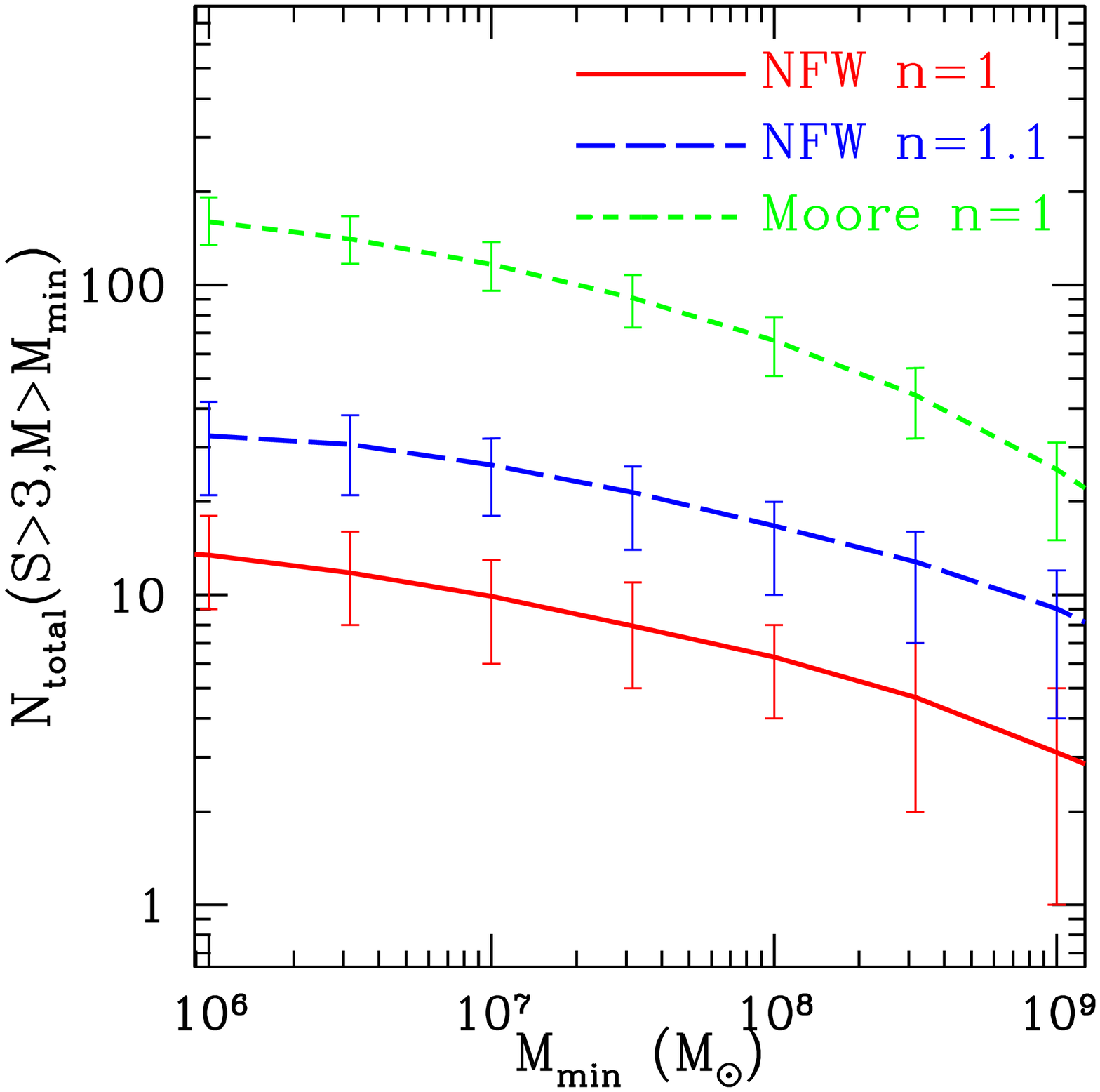,width=6.5cm}
\end{tabular}
\end{center}
\vspace*{0.25cm}
\caption{Left:  The cumulative  number of  subhalos with  mass  $M \ge
\Mmin$ as a function of $\Mmin$ for different values of the 
core parameter $\beta=\tilde{\rc} /\tilde{r}_{\rm  c,  0}$. The  solid,  
long-dashed  and dash-dotted  lines
correspond  to means over  all realizations  in a  $\LCDM$ cosmological
model for $\beta = 10^{-2}$,  $10^7$ and $5 \times 10^7$ respectively.
Error  bars are as  in Figure  \ref{fig:fig1}.  Right:  The cumulative
number of visible subhalos with a  mass $M \ge \Mmin$ for the standard
$\LCDM$ cosmological model  (solid), a model with a  spectral index of
primordial fluctuations $n=1.1$  ($\sigma_8 \simeq 1.2$; long-dashed),
and a  model where the density  profiles of subhalos  are described by
the Moore et al.  profile (short-dashed).}
\end{figure}

A key ingredient in this calculation is the matter distribution in the
very central regions of subhalos.  In the absence of any other effect,
subhalos may  exhibit a central, constant density core established by
the competition  between the rate of neutralino  annihilations and the
rate of infalling material.  Due to the significant uncertainty in the
mass densities achieved  in the central regions of  dark matter halos,
we investigate  the effect of the  size of the core  region on subhalo
detectability.  In Figure \ref{fig:fig2}, we show how our results vary
as a function of core radius, parameterized by $\beta \equiv r_{\rm c}
/r_{\rm c,  0}$, where  $r_{\rm c,0}$ is  the core radius  assigned by
equating the annihilation rate inside the core to the rate of material
infall  (see \cite{KZW03}),  and  $r_{\rm  c}$ is  a  new core  radius
defined as a multiple of  $r_{\rm c,0 }$.  Clearly, the precise choice
of the core radius affects our results only weakly over many orders of
magnitude.  Notably,  making $\beta <  1$ does not have  a significant
effect  on $N_{\rm total}$.   Eventually, when  $\beta \gsim  5 \times
10^7$, the  number of  detectable subhalos decreases  significantly as
the angles subtended by typical subhalo cores become comparable to the
detector resolution.

In general the luminosity and  therefore detectability of a subhalo is
given by  integrating the  square of the  mass density of  the subhalo
along the line-of-sight to the subhalo.  It is of interest to test the
robustness  of our  results  by investigating  the  change in  $N_{\rm
total}$ when  different mass density  profiles are assumed.   For this
purpose,  we show in  Figure \ref{fig:fig2}  the number  of detectable
objects  when  the Moore  et  al.  profile \cite{moore_profile}  (with
$\rho(r) \propto  r^{-3/2}$ at small radii) is  assumed.  As expected,
the number of detectable  subhalos increases dramatically (a factor of
$\sim 10$).


\section{\label{sec:conc}Conclusions}


We investigated the possible detection  of the MW substructure via the
detection  of gamma-rays  from neutralino  annihilations  in otherwise
dark subhalos.  We chose the  most optimistic SUSY parameters in order
to  maximize  the  probability  of  detection.   We  also  employed  a
realistic,  yet still  optimistic  from the  standpoint of  predicting
observable  signals from  substructure,  model for  the population  of
subhalos in the MW.  Our main results were:

$\star$ If  the neutralino  is relatively light  ($\Mchi \lsim  100 \,
\GeV$), then  GLAST and  VERITAS, working in  concert, may be  able to
detect the gamma-ray signal.  In this case, GLAST with its large field
of  view  can be  used  to  identify sources  in  the  sky and  direct
subsequent VERITAS observations, which can search for line-emission at
$E  = M_{\chi}$,  the  smoking gun  of  neutralino annhilations.   For
example,  if $M_{\chi}  \sim 75$  GeV, then,  in the  case  of optimal
coupling to  photons, there  will be $\sim  1$ detectable  subhalo per
GLAST field of  view, on average.  In this  case, subsequent, directed
observations with VERITAS should  be able to confirm the line-emission
feature after an exposure time of $t_{\rm exp} \sim 450$ hr.

$\star$ For neutralino masses in the range $100 \, \GeV \lsim M_{\chi}
\lsim  500$  GeV,  detection  requires  an  instrument  with  a  large
effective area, like  VERITAS; however, such a detection  must rely on
serendipity due to the  small number of potentially detectable objects
and VERITAS' comparably small field of view.

$\star$  For  $\Mchi  \gsim  500  \,  \GeV$  it  seems  unlikely  that
gamma-rays  from neutralino  annihilations  in dark  subhalos will  be
detectable with VERITAS or GLAST.

What can be learned by the lack of such a detection?  The lack of such
a detection certainly will not lead to a bonanza of constraints on the
MSSM  or  SUSY in  general.   Even  after  choosing the  optimal  MSSM
parameters for detection,  the likelihood of a detection  is small for
most  of the  viable range  of  $M_{\chi}$.  Moreover,  the number  of
detectable  objects  depends  upon  the  uncertain  shape  of  density
profiles  in  the  innermost  regions  of dark  matter  halos.   These
uncertainties can significantly  influence our predictions.  Moreover,
their are  additional uncertainties that  are not associated  with our
lack of  knowledge of density  profiles and subhalo  populations.  The
predictions of the expected gamma-ray flux are strongly dependent upon
poorly-constrained    cosmological   parameters.     We    showed   in
Fig. \ref{fig:fig1} that adopting the best-fitting power spectrum from
the WMAP  group reduces the probability  of detection by  more than an
order  of  magnitude  relative  to  a  model  with  a  standard,
scale-invariant primordial power spectrum.

Of course, a detection would yield a great deal of information.  First
and foremost, it would be  evidence for neutralino (or some other WIMP
that annihilates into  photons) dark  matter, it  would  suggest the
presence of dark  subhalos within the MW, it  would indicate that such
subhalos do achieve extremely high densities in their central regions,
and  it  may also  yield  information  about  the survival  rates  and
accretion histories of dark matter subhalos.  Nevertheless, it will be
difficult  to   ``measure''  SUSY  from  such  a   detection.   As  we
demonstrated above,  cosmological parameters play a  role in predicting
the gamma-ray signal and, as  we show in Fig. \ref{fig:fig2}, adopting
a  ``blue tilted''  power  spectrum with  $n=1.1$, COBE-normalized  to
$\sigma_8  \simeq 1.2$, can  boost the  expected number  of detectable
subhalos  by  a  factor  of  $\sim 3$.   Such  uncertainties  must  be
marginalized over  and a model similar  to the one we presented 
here may be able to play an important role in this regard.  Further,  the 
flux from a  particular subhalo depends upon subhalo distance and there is 
no obvious way to determine reliably the distance to an otherwise dark subhalo.  
Our model attempts to take this uncertainty into account by  calculating
``likely'' realizations of the substructure population of the MW.


\vfill 
\end{document}